\pdfoutput=1
\documentclass[useAMS,usenatbib]{mn2e}
\usepackage{graphicx,bm,mathabx,braket,dsfont}

\title[]{On the Evidence for Cosmic Variation of the Fine Structure
  Constant (II): A Semi-Parametric Bayesian Model Selection Analysis of the Quasar Dataset}
\author[E. Cameron and A. N. Pettitt]{E. Cameron$^{1}$\thanks{E-mail:
dr.ewan.cameron@gmail.com} and A. N. Pettitt$^{1}$\\
$^{1}$School of Mathematical
Sciences (Statistical Science), Queensland University of Technology
(QUT), GPO Box 2434,\\ Brisbane 4001, QLD, Australia}
\begin{document}

\date{Submitted to MNRAS: xx July 2013.}

\pagerange{\pageref{firstpage}--\pageref{lastpage}} \pubyear{2013}

\maketitle

\label{firstpage}

\begin{abstract}
In the second paper of this series we extend our 
Bayesian reanalysis of the evidence for a cosmic variation
of the fine structure constant to the  
semi-parametric modelling regime.  By adopting a mixture of Dirichlet processes prior for the
unexplained errors in each instrumental subgroup of the benchmark
quasar dataset we go some way towards freeing our model selection
procedure from the apparent subjectivity of a fixed distributional
form.  Despite the infinite-dimensional domain of the
error hierarchy so constructed we are able to demonstrate a recursive scheme for marginal likelihood estimation with
prior-sensitivity analysis directly analogous to that presented in
Paper I, thereby allowing the robustness of our posterior Bayes
factors to hyper-parameter choice and model specification to be
readily verified.  In the course of this work we elucidate various 
 similarities between unexplained error problems in the seemingly disparate fields of astronomy and clinical meta-analysis, and we highlight a number of sophisticated techniques for handling such problems made available by past research in the latter.  It is our hope that the novel approach to semi-parametric model selection demonstrated herein may serve as a useful reference for others exploring this potentially difficult class of error model.
\end{abstract}


\section{Introduction}\label{introduction}
Recent claims by \citet{web11} and \citet{kin12} of cosmic variation in the fine structure constant,
$\alpha$, appearing to the Earth-bound observer as a North-South
dipole have generated considerable
interest owing to the remarkable
theoretical implications of such a result, if true.  As acknowledged by
the Webb et al.\ team  though, the quasar dataset
upon which these claims are based exhibits clear signs of an
unexplained error term, $\varepsilon_\mathrm{sys}$, manifest as a
residual variance over and above the level of noise attributable to
known sources of observational uncertainty; the
presence of which confounds the statistical interpretation of any
alleged spatial trends.  Importantly, a skeptical
reader might well disagree with the assumptions of a strictly unbiased
and Normally-distributed $\varepsilon_\mathrm{sys}$ used by the Webb
et al.\ team to derive the
quoted ``$4\sigma$ significance'' of their dipole hypothesis.  

In Paper I \citep{cam13a} we highlighted the appearance of multimodality and/or
heavy-tailedness in the distributions of residuals about the best-fit
parameterization of the proposed dipole model and explored a variety
of \textit{parametric} forms for the unexplained error term in a Bayesian model selection (BMS)
framework.  Our study revealed weak support for a scenario
in which $\alpha$ is truly constant but the Keck and VLT observational subgroups of the quasar dataset
are subject to non-zero mean/mode noise of opposing sign.  Moreover,
despite the ``Ockham's Razor''-like tendency of BMS to favour simpler
models our Bayes factor comparison in Paper I also gave support to a proposed skew Normal
parent distribution for the $\varepsilon_\mathrm{sys}$ term.  In the interests of thoroughness we herein
refine our modelling approach (and thereby relax further our minimal
assumptions) to admit an ``infinitely
flexible'', non-parametric
representation of these unexplained uncertainties.  The mathematical basis of our new
approach is the mixture of Dirichlet processes prior.

Introduced by
\citet{fer73}, the Dirichlet process (DP) represents an
infinite-dimensional extension of the ordinary Dirichlet distribution;
the importance of which for applied statistics lies in the fact that
its topological
support includes a weak approximation to every probability distribution on
a given domain.  This property of the DP---and its hierarchical
extension, the mixture of Dirichlet processes (MDP) prior \citep{ant74}---has allowed the
construction of novel solutions to 
a wide variety of challenging inference problems featuring some
element of non-parametric density estimation, most notably within
the mixture modelling framework  
\citep{esc95,mac98,nea00}.  A number
of recent astronomical applications have evinced the potential of DP-
and MDP-based analyses in this field as well: e.g.\ for the classification of gamma ray bursts according
to their
observed spectral slopes \citep{cha07}, the identification of
variability in irregularly-spaced time series \citep{shi09}, and the estimation of galactic
potential shapes from detailed kinematical data \citep{mag13}.

While there exist a number of computationally efficient, publically
available packages for sampling from the  posteriors of
basic DP and MDP models\footnote{E.g.\ the \texttt{bspmma} \citep{bur12} and \texttt{DPpackage}
\citep{jar07} add-on libraries in
\texttt{R}.}, the challenge lies in adapting these packages for the
solution of real-world, bespoke analysis problems, especially in the BMS context.  In
this study we explicitly demonstrate one such adaptation procedure, in which we highlight the utility
of the recursive approach to marginal likelihood
estimation under an MDP prior; 
to this end we make particular use of the Radon-Nikodym derivative for the
DP derived by \citet{dos09}.  It is our hope that the novel approach to semi-parametric model selection exhibited herein may serve as a useful reference for other astronomical studies exploring this potentially difficult class of error model.

The structure of this paper is as follows.  In Section \ref{dpmm} we
briefly review key properties of the DP, before
formulating our precise hierarchical model for the quasar dataset and
specifying priors for its controlling parameters.  In
Section \ref{meta} we elucidate key similarities between the
unexplained error problems faced in astronomical and clinical
meta-analysis studies with the aim of facilitating cross-disciplinary
learning in this regard.  In
Section \ref{rlr} we describe a procedure for marginal likelihood
estimation under our semi-parametric model, and in Section \ref{isr}
we detail a corresponding importance sample reweighting scheme for
prior-sensitivity analysis.  In Section \ref{results} we present the
results of this BMS methodology applied to the quasar dataset and discuss
their interpretation in light of our earlier parametric modelling.  Finally,
in Section \ref{conclusions} we summarize our conclusions.

\section{Dirichlet Process Mixture Model}\label{dpmm}
\subsection{The Dirichlet Process}\label{dp}
The DP \citep{fer73} defines a stochastic process of random probability
measures on the atomic elements in the 
 $\sigma$-algebra, $\Sigma_\Omega$, of a sample space, $\Omega$.  As
 such the DP may be seen as an infinite-dimensional
extension of the finite-dimensional Dirichlet
distribution (familiar to Bayesians as the conjugate prior for a
multinomial likelihood function).  Under its defining 
 parameter pairing of concentration index, $M > 0$,
and (normalised) centering distribution, 
 $G$, on
 $\left<\Sigma_\Omega,\Omega\right>$ the DP (as a random measure, $P$) may be characterised by the
 following property: that for every (measurable) partition, $\{B_1,\ldots,B_k\}$,
 of $\Omega$ (and for all $k=1,2,\ldots$) the distribution of
 $[P(B_1),\ldots,P(B_k)]$ is (finite-dimensional) Dirichlet with parameter vector,
 $[M\times G(B_1),\ldots,M\times G(B_k)]$.  Proof that the above is
 sufficient to define a unique stochastic process
 follows ultimately from Kolmogorov's existence theorem (cf.\ \citealt{fer73}).
 
Key properties of the DP include:
 \textsc{(i)} that, although strictly atomic, its topological
 support\footnote{In the topology of pointwise convergence \citep{fer73}.}
 is in fact the complete set of probability measures on
 $\left<\Sigma_\Omega,\Omega\right>$ absolutely continuous with
 respect to (and with supports enclosed by that of) the centering distribution, $G$; \textsc{(ii)}
 realisations of (i.e., draws of 
 probability distributions, $P$, from) the DP 
 may be simulated via a
 ``stick-breaking'' \citep{set94} construction, while its marginal
 output (i.e., iid draws from $P$ marginalised over $\mathrm{DP}(M,G)$) may be simulated via a P{\'o}lya urn \citep{bla73} scheme; \textsc{(iii)}
 owing to the discreteness of $P \sim \mathrm{DP}(M,G)$, individual
 values drawn from any one such $P$ will
 eventually repeat themselves (almost surely); \textsc{(iv)} the
 expectation of $P(A)$ with respect to the DP with centering distribution,
 $G$, is simply $G(A)$; and \textsc{(v)} the expected mean and
 variance of $P \sim \mathrm{DP}(M,G)$ are $\mathrm{E}\{G\}$ and $\frac{M}{M+1}\mathrm{Var}\{G\}$.  Proofs of \textsc{(i)}, \textsc{(ii)},
  \textsc{(iv)}, and \textsc{(v)} are given by \citet{fer73} (and those for
 \textsc{(iii)} are as cited above); in addition, \citet{gho10} offers a handy
 survey of these and other relevant insights.

\subsection{Hierarchical Dirichlet Process Models}\label{hdpm}
The ``infinite flexibility'' of the DP, described by property
\textsc{(i)} above, has led to its widespread use as a non-parametric
modelling device for Bayesian inference \citep{fer83}.  A particularly
well-studied construction of this sort is represented by the
following hierarchical model (cf.\ \citealt{lo84}, \citealt{esc95},
\citealt{bur05}):
\begin{eqnarray}
y_{i}|\mu_i,x_i &\sim& f(\cdot|\mu_i,x_i), \ \ \ i=1,\ldots,n \\
\mu_1,\ldots,\mu_n|P &\sim& P,\\
P|M,\psi &\sim& \mathrm{DP}(M,G_\psi),\label{hdp}\\
M,\psi &\sim& F.
\end{eqnarray}
That is, each observed datapoint, $y_i$, for $i=1,\ldots,n$, is assumed
drawn from a parametric distribution with density,
$f(\cdot|\mu_i,x_i)$, defined by two controlling parameters, $x_i$ and
$\mu_i$, with the former known exactly and the latter a hidden sample
from a hidden
realization, $P$, of the Dirichlet process, DP$(M,G_\psi)$.  For generality,
 the concentration index, $M$, of this DP, and the controlling
 parameters, $\psi$, of
 its centering distribution, $G_\psi$, are also shown here as random
 elements, drawn from some (parametric) density, $F$.  Such a model can be described as
\textit{semi-parametric} (e.g.\ \citealt{bas03}) since it combines a
non-parametric form for the $\mu_i$ with a parametric form for each 
$y_i|\mu_i,x_i$.  Often $f(\cdot|\mu_i,x_i)$ will be chosen as
the univariate Normal with mean,
$\mu_i$, and standard deviation, $x_i$; in which case the above may
also be referred to as a  non-parametric ``random effects'' model (especially in the
clinical meta-analysis
setting).

In the spirit of our earlier \textit{parametric} modelling of the
fine-structure dataset (Paper I) we adopt here for the total error term,
$\varepsilon_\mathrm{tot}$, operating on the $\Delta \alpha/\alpha$
estimates of the Webb et al.\ team's quasar
dataset the following 
specific version of this classic
semi-parametric model:
\begin{eqnarray}
\label{block} \varepsilon_{\mathrm{tot},i}|\mu_{\mathrm{sys},i},\sigma_{\mathrm{ran},i}
& \sim &\mathcal{N}(\mu_{\mathrm{sys},i},\sigma_{\mathrm{ran},i}^2),\ i=1,\ldots,n_g \\
\mu_{\mathrm{sys},1},\ldots,\mu_{\mathrm{sys},n_g}|P &\sim&
P \\
P | M,  \sigma_\mathrm{sys} &\sim&
\mathrm{DP}(M,\mathcal{N}(0,\sigma_\mathrm{sys}^2))\\
\sigma_\mathrm{sys} | M &\sim&
\mathcal{N}_\mathrm{half}(0, \sigma_p^2 \times \frac{M+1}{M}) \label{end} \\
M &\sim& \Gamma(\gamma_1,\gamma_2). \label{Mdist}
\end{eqnarray}
Here $\mathcal{N}(\mu,\sigma^2)$ denotes the Normal with mean, $\mu$,
and standard deviation, $\sigma$, and 
$\Gamma(\gamma_a,\gamma_b)$ denotes the Gamma distribution with shape,
$\gamma_a$, and rate, $\gamma_b$.  Important to note is that we allow
a unique $M$, $\sigma_\mathrm{sys}^2$, $P$, and
$\{\mu_{\mathrm{sys},1},\ldots,\mu_{\mathrm{sys},n_g}\}$ for each of
the three instrumental subgroups (Keck LC, Keck HC, and VLT) of the
quasar dataset; hence, the $n_g$ in the above refers to the number of absorbers
detected in each subgroup (i.e., 113, 27, and 153, respectively)
rather than the total sample size (293).  

As described in Paper I, the use of a Normal form
for the explained uncertainty term appearing in the top layer of the above
hierarchy follows naturally from the known details of the 
Webb et al.\ team's $\Delta \alpha/\alpha$ fitting procedure \citep{kin10}.  Our choice of a Normal centering distribution for
the Dirichlet process on the third layer is, in contrast, purely a
matter of computational convenience, allowing for efficient Gibbs
sampling from the resulting MDP posterior given conjugate hyperpriors on
$\sigma_\mathrm{sys}$ and $M$ (as we discuss further below).  The sensitivity of the resulting Bayes factors to
this particular modelling decision can, however, be readily explored via the importance
sample reweighting scheme described in Section \ref{isr}; and indeed
we present the results of such an analysis, supposing instead a Student's $t$
centering distribution for our DP, in Section \ref{results}.  It is
also  
worth observing here that although our chosen centering distribution
is strictly zero mean (and mode) the same does not follow for any
particular draw, $P$, from this DP.  In the terminology of Paper I
this is a \textit{biased} error model.  

As mentioned above, given appropriate hyperpriors on
$\sigma_\mathrm{sys}$ and $M$, the posterior of such a semi-parametric
error model can be efficiently explored via a Gibbs sampling
scheme. The mathematical and algorithmic details of this procedure are already well-described by
\citet{esc95} and \citet{nea00} so we omit these from the present
account; although we return to the topic of Gibbs sampling again in
Section \ref{gibbs}.  The conjugate priors required in this case are
independent Gamma densities on $M$ and $1/\sigma_\mathrm{sys}^2$; the former we
have already in Equation \ref{Mdist}, but the latter we do not.  In
fact, our prior on $\sigma_\mathrm{sys}$ is defined as a half Normal
conditional on $M$, chosen so as to achieve a degree of consistency
with our prior on the equivalent component (namely, the standard deviation) of the parametric
unexplained error terms considered in Paper I.  The conditioning on
$M$ is designed to match $\sigma_p^2$ with the expected variance
of $P$ as per Property \textsc{(v)} of the DP given in
Section \ref{dp}.  

In order to reconcile this non-conjugate prior
choice with the power of efficient Gibbs sampling under  conjugate
priors we use the following trick.  First, we replace our
nominal prior on $\sigma_\mathrm{sys}|M$ with the conjugate prior form on
$1/\sigma_\mathrm{sys}^2$---namely, $1/\sigma_\mathrm{sys}^2 \sim
\Gamma(\gamma_3,\gamma_4)$---with $\gamma_3$ and $\gamma_4$ chosen (by
eye) for similarity with our true prior
(marginalized over $M$). After exploring the resulting posterior (and
its tempered
bridging densities) from this conjugate prior (Section \ref{gibbs}) and estimating marginal
likelihoods accordingly (Section \ref{rlr}), we finally use importance
sample reweighting (Section \ref{isr}) to
recover the true posteriors and marginal likelihoods under our true (nominal)
prior.  For reference, our default hyperparameter choices are $\gamma_1 = 1.5
$, $\gamma_2 = 0.05$, and $\sigma_p = 2\times 10^{-5}$, for which we
find $\gamma_3=0.4$ and
$\gamma_4=0.04$ to give a reasonably well-matched conjugate form.

\subsection{Hypotheses}\label{hyp}
To specify a complete generative model for the quasar dataset we couple the
above semi-parametric error model to one of three (parametric) hypotheses for cosmic $\alpha$
variation: \textsc{(i)} the null hypothesis (that the fine structure
constant is everywhere exactly equal);
\textsc{(ii)} the monopole hypothesis (that of a fixed cosmic offset
relative to its laboratory value); and \textsc{(iii)} the
monopole$+$$r(z)$-dipole hypothesis (that of a large-scale cosmic spatial trend).  In particular, we adopt the
following form for this dipole after \citet{kin12}:
\begin{equation}\label{dipoleeqn}
\Delta \alpha /
\alpha{}_{\mathrm{mod}|\bm{x}_i,\bm{\theta}_m} = m + B \times r(z_i)
\cos (\phi)[\mathrm{ra}_i,\mathrm{dec}_i,\mathrm{ra}_d,\mathrm{dec}_d]
\end{equation}
with $\bm{x}_i=\{\mathrm{ra}_i,\mathrm{dec}_i,r(z_i)\}$ a vector of
explanatory variables for the $i$th
absorber, $\bm{\theta}_m=\{
m,B,\mathrm{ra}_d,\mathrm{dec}_d\}$
a vector of input model parameters, and $\cos (\phi)[\cdot]$ a
function returning the cosine of angular
separation between the observational sightline and dipole
vector.  Formulae for $\cos (\phi)[\cdot]$ and $r(z)$ are given in
Paper I.  In this notation the monopole and strict null hypotheses
become simply $\Delta \alpha /
\alpha{}_{\mathrm{mod}|\bm{\theta}_m} =m$ and $\Delta \alpha /
\alpha{}_{\mathrm{mod}} = 0$, respectively.  As in Paper I we suppose the following priors on the parameters of
these hypotheses:
\begin{eqnarray}
m &\sim& \mathcal{N}(0,[0.5\times10^{-5}]^2)\\
B &\sim& \mathrm{Exp}(1/[0.5\times10^{-5}])\\
\mathrm{ra}_d &\sim& \mathcal{U}(0,24)\\
\sin \mathrm{dec}_d &\sim& \mathcal{U}(-1,1).
\end{eqnarray}

Finally, we note that the resulting likelihood function
conditional on the collection of hidden $\{\bm{\mu}_\mathrm{sys}\}=\{\{\mu_{\mathrm{sys}}\}_{(1)},\{\mu_{\mathrm{sys}}\}_{(2)},\{\mu_{\mathrm{sys}}\}_{(3)}\}$ from each
instrumental group takes the simple form: \begin{eqnarray} \nonumber
&L(\bm{y}|\bm{\theta}_m,\{\bm{\mu}_{\mathrm{sys}}\},\{\bm{x}\}) =& \\
&\prod_{i=1}^{293}
f_{\mathcal{N}(\mu_{\mathrm{sys},i},\sigma_{\mathrm{ran}_i}^2)}(\Delta\alpha/\alpha{}_i -
\Delta\alpha/\alpha{}_{\mathrm{mod}|\bm{x}_i,\bm{\theta}_m}).\label{likelihood}&
\end{eqnarray}

\subsection{Gibbs Sampling from the Full Posterior}\label{gibbs}
As mentioned earlier, particular aspects of our hierachical error model
were chosen to facilitate Gibbs sampling of the error model posterior via the
P{\'o}lya urn technique popularized by \citet{esc95} and others.  The Gibbs sampler (cf.\ \citealt{cas92}) is a commonly used procedure
for Markov Chain Monte Carlo (MCMC) exploration of multi-variate posteriors
by way of simulation from only some lower-order conditionals; the 
popularity of which has surged in recent years
in concert with the popularity of Bayesian hierarchical modelling.  
Past astronomical applications of the Gibbs sampler include procedures for supernova
light-curve fitting \citep{man09} and spectral analysis in the
low-count limit \citep{van01}.

To explore the joint posterior of a given hypothesis plus
semi-parametric error model pairing we adopt a two-part
Gibbs sampling approach in which the hypothesis parameters are updated
conditional on the current error model parameters, and then these current error model
parameters are updated conditional upon the new hypothesis parameters.
That is, we draw $\bm{\theta}_m^{(i)} \sim
\pi(\bm{\theta}_m|\{\bm{\mu}_{\mathrm{sys}}\}^{(i)})$ and
$\{\bm{\mu}_{\mathrm{sys}}\}^{(i+1)} \sim 
\pi(\{\bm{\mu}_{\mathrm{sys}}\}|\bm{\theta}_m^{(i)})$.  For the former we run
a simple random walk MCMC chain for 100 moves (to approximate stationarity) under the conditional
likelihood function given by Equation \ref{likelihood}, and for the
latter we run the \texttt{DPmeta}
routine of the \texttt{DPpackage} in \texttt{R} \citep{jar07} for 1000
moves from the current state \textit{for each of the three
  instrumental subgroups of the quasar dataset} (separately).  Note
that the second stage here implicitly updates the current $M$ and
$\sigma_\mathrm{sys}$ for each error group as well, and the
\texttt{DPmeta} routine itself performs the Gibbs sampling scheme of
\citet{esc95} under our conjugate prior proxy (cf.\ Section \ref{hdpm}).  

An important observation for the purposes of our subsequent marginal likelihood
estimation is that the
tempered likelihood posterior may also be readily explored via the
above algorithm.  In particular, to simulate from the tempered
posterior at temperature, $\beta$, we need simply replace the 
$\sigma_i$ in Equations \ref{block} and \ref{likelihood} with $\sigma_i/\sqrt{\beta}$,
which gives the tempered likelihood up to a constant of proportionality
(the value of which is irrelevant for the present application).
Before detailing the recursive scheme used to this
end in Section \ref{rlr} we first review a number of similiarities
between the hierarchical, semi-parametric model outlined here and the
equivalent random effects model from clinical meta-analysis in Section
\ref{meta} below.  The purpose of this digression is to better qualify
the place of our model in a wider applied statistics context and
perhaps therefore to encourage some beneficial cross-disciplinary learning in
this regard.

\section{Unexplained Error Problems in Astronomy and Clinical
  Meta-Analysis }\label{meta}
Unexplained error problems, typically invoked \textit{a posteriori} when the residuals about some well-trusted regression
model are found to far exceed the explained variance of the
measurement process, seem to arise
quite regularly in the practise of astronomical data
analysis.  For instance, the Sunyaev-Zel'dovich effect-based cluster dataset compiled by
\citet{gal13} for probing cosmic $\alpha$ variation exhibits a
unexplained error term just as problematic at that for the quasar dataset
studied here.  Despite this, the statistical treatment of such unexplained errors in astronomical
studies remains rather ad-hoc, and, against warnings
\citep{and10}, many astronomers still favour the na\"ive approach of
simply re-scaling their explained uncertainty estimates to enforce a
reduced-$\chi^2$ of one.  In contrast, the techniques now
considered routine for handling the equivalent random effects problem in the field of 
clinical meta-analysis, in which the published treatment effect estimates from
 multiple trials of the same intervention are combined for enhanced
 statistical power, are generally far more sophisticated. 

In a standard meta-analysis each study contributes its own unique 
estimates of both the targetted treatment effect, $\hat{\phi}_i$, and
the inherent 
variance, $\hat{\sigma}_i^2$, 
of this measurement according to its explained (or ``within-study'')
error term.  The distributional form of the latter 
 may often be assumed Normal from theoretical
 considerations of the known sampling design and inference procedure
 (e.g.\ maximum-likelihood); and the further simplifying assumption of
 $\sigma_i^2 = \hat{\sigma}_i^2$ (i.e., treating the estimated
 variance as if it were the known truth) yields the first level of a
 hierarchical model for each observed datapoint, $\hat{\phi}_i
 \sim \mathcal{N}(\phi_i,\sigma_i^2)$.  (Note that in effect we have
 made the same assumption implicitly in our own error model of Section
 \ref{hdpm}.)  The unexplained (or ``between-study'')
 error term over the complete dataset may then be modelled on a
 second level under the strong assumption of a known
 distributional form with unknown variance, $\Sigma^2$, and unknown centering
 parameter, $\mu$; the latter defining \textit{the} reference effect
 being targetted
 in the 
 meta-analysis paradigm.  

Under a heuristic invocation of the Central Limit Theorem, and perhaps ultimately for
 computational convenience, a Normal 
 distribution is typically supposed for this purpose.  In this case
 the resulting 
 (parametric) hierarchical model may be written as follows:
\begin{eqnarray}
\hat{\phi_i} \sim \mathcal{N}(\phi_i,\sigma_i^2)\\
\phi_i \sim \mathcal{N}(\mu,\Sigma^2).
\end{eqnarray}
Upon the specification of appropriate priors for $\mu$ and
$\Sigma$ a computational exploration of the corresponding posterior
may be easily conducted, allowing prediction of the observed
treatment effect(s) from future studies (cf.\ \citealt{der86,hig09}). 

A desire to limit the potential for subjective bias arising from the assumption
of a particular distributional form for the unexplained error term in this
context has led to the development of various semi-parametric meta-analysis schemes.
\citet{bur05} give a popular formulation almost identical in structure to the
hierarchical error model used here for the quasar dataset; the key
difference being that a \textit{conditional}
Dirichlet process with explicitly controlled median is used in their study in place of the
ordinary Dirichlet process to clarify the interpretation of inference
on $\mu$.  Code for posterior simulation from the \citet{bur05} model
is available in the  \texttt{bspmma} package for \texttt{R}.  Although
coded under the assumption of a fixed concentration index for the
conditional Dirichlet, the posteriors for a series of \texttt{bspmma}
runs over a suitable range of
$M$ can ultimately be
combined to represent sampling from a prior on this parameter via the
recursive technique, as per \citet{dos09} (which we will discuss in more
detail later).

For meta-analysis problems in which the studies compiled into the
dataset divide naturally amongst distinct error groups---just like the
case of the Webb et al.\ team's quasar dataset considered here---\citet{mue04}
outline a novel scheme for enhanced learning of the error
distribution across groups.  As a compromise between the archetypal modelling 
extremes of completely non-interacting error groups (i.e., each takes its own $M$ and $G_\psi$) and strongly interacting
error groups (i.e., all share a common $M$ and $G_\psi$) the
\citet{mue04} model retains the structure of the former but adds a new
shared error layer inducing a dependence similar to the latter.
  The utility of this technique for powering up inference in the small
  sample size regime ($\sim$50 members per group) is illustrated by
  \citet{mue04} in application to a challenging cancer research
  dataset.  Although we prefer the simplicity of a non-interacting model for our
  present analysis (in which we also have somewhat larger groups) it is
  worth noting that an efficient
Gibbs sampling algorithm for posterior exploration under the shared layer error
model is also available in the \texttt{DPpackage} for \texttt{R}.  

\section{Recursive Marginal Likelihood Estimation}\label{rlr}
Our aim here, as in Paper I, is to evaluate the relative
strengths of the Webb et al.\ team's three hypotheses (Section \ref{hyp}) for the spatial variation of
$\alpha$ (or lack thereof) in accordance with the BMS paradigm.
Although estimation of the required Bayes factor (the ratio of marginal
likelihoods under two competing hypotheses) is, for fully parametric models, a
very well-studied field of statistical theory  (see, e.g.,
\citealt{fri12} for a recent review) the same is  not true
for the semi-parametric regime.  The additional perceived difficulty in this
case being the infinite-dimensional structure of the prior domain,
which cannot be characterized as a probability density with respect
to the ordinary Lebesgue measure.  In fact, to-date only \citet{bas03}
have explicitly considered this problem in depth; these authors
developing an extension of the popular Chib marginal
likelihood estimator \citep{chi95} suitable for application to semi-parametric
Dirichlet process problems in clinical meta-analysis in which the normalized conditional posterior,
$\pi(\bm{\theta}_m|\bm{y},\{\phi\})$, 
is analytically tractable.  

Here we show how the recursive estimator
resulting from the biased sampling theory of \citet{var85} (see also
\citealt{gil88} and 
\citealt{kon03}) gives a more general means of marginal likelihood
estimation for such semi-parametric problems, requiring only the
availability of a collection of conditional likelihoods (Equation
\ref{likelihood}) drawn in known proportions, $n_1/n, \ldots, n_m/n$
with $n = \sum_{i=1}^m n_i$, from a sequence of tempered
posteriors at known temperatures, $\beta_1,\ldots,\beta_m$, with $\beta_1=0$
(the prior; $Z_1=1$) and $\beta_m=1$ (the full posterior; $Z=Z_m$).  The first step is
to observe that the prior for our hierarchical model defines a proper
probability measure, $P$, on the Borel sets of the metric
space, $S$, corresponding to
the product of the domains of $\{\bm{\mu}_\mathrm{sys}\}$,
$\bm{\theta}_m$, and $\bm{T}$---with
$\bm{T}=\{\bm{\tau}_{(1)},\bm{\tau}_{(2)},\bm{\tau}_{(3)}\}$ explicitly
representing the collection of $\bm{\tau}=\{M,\sigma_\mathrm{sys}\}$
pairs for each of our three error groups.
We then observe that the conditional likelihood function defines a
continuous mapping from $S$ to another metric space, $S^\prime$, being
simply the real line.  Hence, by the transformation of variables
theorem we know that $P$ induces a proper probability measure,
$P^\prime$, on $S^\prime$ with $Z_1 = \int_{0}^\infty dP^\prime(l)
= 1$.  The expectation under this new measure is
simply the marginal likelihood itself, $Z=\int_{0}^\infty l\ 
dP^\prime(l)$, and likewise for the normalizations of the tempered posteriors, $Z_\beta=\int_{0}^\infty l^\beta
dP^\prime(l)$.  That is, despite the complexities of our prior the
continuous mapping provided by the likelihood function returns us to
the canonical biased sampling framework.  

Although we do not know $P^\prime$ the non-parametric maximum
likelihood estimator given by \citet{var85} nevertheless allows for unbiased
estimation of the associated $Z_{k}$ ($k \neq 1$).  (Important to note
though is that only the updated convergence analysis given by
\citealt{gil88} confirms the validity and well-behavedness of this
scheme for non-Lebesgue densities and indeed for general sample spaces.)  The recursive relation
so defined takes the same form as that for reverse logistic regression
under tempered posterior exploration given in Paper I.  Namely, 
\begin{equation}
\hat{Z}_{k} = \sum_{i=1}^n \left( L_i^{\beta_k} / [ \sum_{j=1}^m
  n_j L_i^{\beta_j} / \hat{Z}_j] \right)
.\end{equation}
Importantly, iteration over this system of equations yields a globally
convergent solution, easily recovered computationally.  Of course, as
noted in Section \ref{hdpm}, our posterior exploration is actually
performed under a conjugate prior proxy for our true (non-conjugate)
priors, meaning that an additional importance sample reweighting step
is now needed to modulate the above, as we describe in detail below.  

\section{Importance Sample Reweighting via the Radon-Nikodym
  Derivative}\label{isr}
As emphasised in Paper I (and also in \citealt{cam13a}), one of the
 strengths of the combination of tempered posterior exploration
with the recursive pathway to marginal likelihood
estimation is that it greatly facilitates the process of
prior-sensitivity analysis.  With the posterior Bayes factor
characteristically sensitive to prior specification this step should
be considered a routine part of all BMS studies \citep{kas95},
although here it is also necessary for recovery of our true posteriors
from the conjugate prior proxies adopted for efficient Gibbs sampling
of our semi-parametric error term (Section \ref{hdpm}).

However, unlike in Paper I for which our priors all admitted probability
densities with respect to Lebesgue measure, the DP
component here clearly does not---as \citet{dos09} observes, there is
a non-zero probability that some of the $\{\bm{\mu}_\mathrm{sys}\}$ will be identical.
For this reason the importance sample reweighting formula needs to be
written in terms of the Radon-Nikodym derivative with respect to our
original prior measure, which we will again denote, $P$.  That is, if
we consider the pool of our tempered posterior draws of
$\{\bm{\theta}_m,\{\bm{\mu}_\mathrm{sys}\},\bm{T}\}$ triples as belonging to the
(pseudo-)importance sampling proposal with probability measure, \begin{equation}H(\bm{\theta}_m,\{\bm{\mu}_{\mathrm{sys}}\},\bm{T}) =
\left(\sum_{i=1}^m n_i/n
L_i^{\beta_i}/\hat{Z}_{i}\right)
P(\bm{\theta}_m,\{\bm{\mu}_{\mathrm{sys}}\},\bm{T}),\end{equation}
then given the Radon-Nikodym derviative of the alternative prior,
$P_\mathrm{alt}$, with respect to this proposal,
$\frac{dP_\mathrm{alt}}{dH}(\bm{\theta}_m,\{\bm{\mu}_{\mathrm{sys}}\},\bm{T})$, we
recover the importance sample reweighting estimator of the marginal
likelihood under $P_\mathrm{alt}$, 
\begin{equation}
\hat{Z}_\mathrm{alt} = 1/n \sum_{i=1}^n L_i \frac{dP_\mathrm{alt}}{dH}(\{\bm{\theta}_m,\{\bm{\mu}_{\mathrm{sys}}\},\bm{T}\}_i).
\end{equation}

\begin{figure*}
\vspace{0pc}
\begin{center}
\includegraphics[width=0.825 \textwidth]{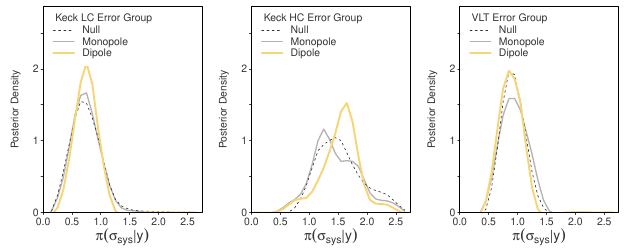} \end{center}
\vspace{-0.275cm}\caption{Illustration of our posterior densities for
  the $\sigma_\mathrm{sys}$ parameter controlling the standard deviation
  of the Normal centering distribution for the DP in our semi-parametric error
  model.  In each panel we compare the resulting posterior under each
  of the null, monopole, and dipole hypotheses for the spatial variation
  (or lack thereof) in $\alpha$; and the ordering (as labelled) of the
three error groups in these panels from left to right is Keck LC, Keck
HC, and VLT.  Consistent with the Webb et al.\ team's original
(non-Bayesian) analysis, our $\sigma_\mathrm{sys}$
 posteriors indicate a far greater degree of 
 unexplained variance in the Keck HC error group than in the Keck LC
 or VLT error groups.}
\label{fig1}
\end{figure*}

\begin{figure*}
\vspace{0pc}
\begin{center}
\includegraphics[width=0.825 \textwidth]{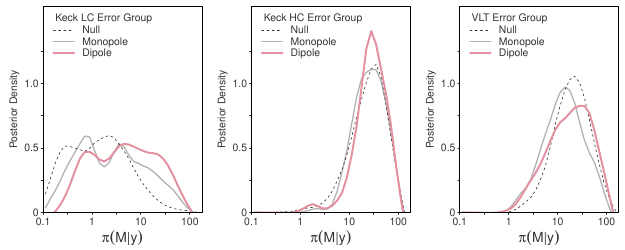} \end{center}
\vspace{-0.275cm}\caption{Illustration of our posterior densities for
  the $M$ parameter controlling the concentration index for the DP in our semi-parametric error
  model.  Evidently the Keck HC and VLT unexplained errors are
 somewhat closer to Normal than those for the Keck LC error group; the
 preference for low $M$ values here suggests that a fixed offset or
 bias would be sufficient to explain the residuals in this group.}
\label{fig2}
\end{figure*}

Importantly, the Radon-Nikodym derivative for the standard
MDP prior model of Equations \ref{block} to \ref{Mdist}---which we
shall denote, $\frac{d\nu_\mathrm{alt}}{d\nu_{H}}(\{\mu_\mathrm{sys}\}_{(g)},\tau_{(g)})$---has already been
derived by \citet{dos09}.  The exact form of this expression is not
reproduced here, but may be
seen in his Equation 2.2.  Extending this formulation over the three
error groups of the quasar dataset plus the space of hypothesis priors
is trivial given their simple product space structure, i.e.,
\begin{equation}
\frac{dP_\mathrm{alt}}{dH}(\bm{\theta}_m,\{\bm{\mu}_{\mathrm{sys}}\},\bm{T})
=
\frac{\pi_\mathrm{alt}(\bm{\theta}_m)}{\pi(\bm{\theta}_m)} \times \prod_{i=1}^3 \frac{d\nu_\mathrm{alt}}{d\nu_{H}}(\{\mu_\mathrm{sys}\}_{(i)},\bm{\tau}_{(i)}).
\end{equation}

An interesting connection between the recursive method for marginal
likelihood estimation and the importance sample reweighting scheme
presented here appears by way of the reference to \citet{dos09}.  In
this earlier work (and see also \citealt{bur05}) the same recursive
estimator was invoked to facilitate the computation of relative Bayes
factors under fixed $M$ semi-parametric models, and against entirely
parametric models; although it seems that the final step of absolute
marginal likelihood estimation for competing hypotheses tied to this
model (as above) was not considered.

\section{Results and Discussion}\label{results}
\subsection{Posteriors}

\begin{figure*}
\vspace{0pc}
\begin{center}
\includegraphics[width=0.825 \textwidth]{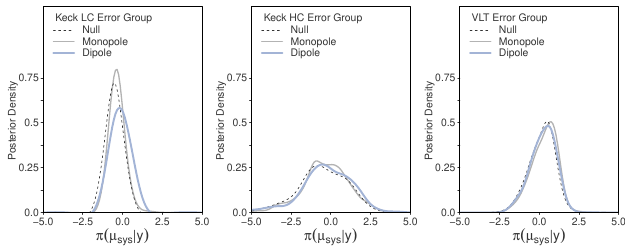} \end{center}
\vspace{-0.275cm}\caption{Illustration of our posterior densities for
  the pooled $\{\mu_\mathrm{sys}\}$ at the top layer of the hierarchy in our semi-parametric error
  model.    The similarity in the distribution of
 $\{\mu_\mathrm{sys}\}$ between hypotheses is
 quite remarkable; the only exception being in the Keck LC
 error group where the dipole hypothesis contributes an effect
 indistinguishable from that of a simple 
 bias term.  }
\label{fig3}
\end{figure*}

\begin{figure*}
\vspace{0pc}
\begin{center}
\includegraphics[width=0.825 \textwidth]{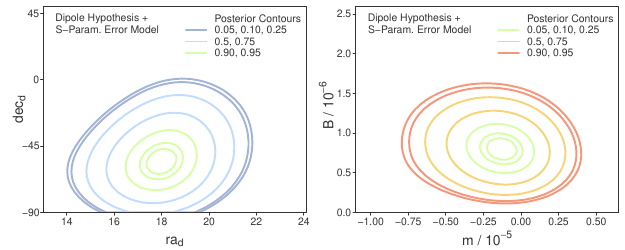} \end{center}
\vspace{-0.275cm}\caption{Illustration of our joint posteriors for
  the $\mathrm{ra}_d$--$\mathrm{dec}_d$ (right ascension--declination of
  the dipole axis) and $m$--$B$ (monopole--dipole component strength) 
   parameter
  pairings for the dipole hypothesis under our semi-parametric error
  model.  The (smoothed) contours bound equi-density regions enclosing 5, 10,
  25, 50, 75, 90, and 95\% of posterior mass (as indicated). These
  hypothesis parameter posteriors are in fair agreement with those
  recovered under the biased parameteric models explored in Paper I.}
\label{fig4}
\end{figure*}

Although only the marginal likelihoods (Section \ref{mls}) of each hypothesis for spatial
variation (or lack thereof) in the fine structure constant are required for our BMS analysis the
posterior parameter distributions explored here offer an important guide to
 the nature of the semi-parametric error term.  In Figures
 \ref{fig1} and \ref{fig2} we compare the posteriors for
 $\sigma_\mathrm{sys}$ (the standard deviation of our Normal centering
 distribution for the DP) and $M$ (its concentration parameter) in
 each of the three instrumental subgroups of the quasar dataset (Keck low/high
 contrast, and VLT) under each of the null, monopole, and dipole
 hypotheses.  Consistent with the Webb et al.\ team's original
 (non-Bayesian) analysis our $\sigma_\mathrm{sys}$
 posteriors indicate a far greater degree of 
 unexplained variance in the Keck HC error group than in the Keck LC
 or VLT error groups.  While from inspection of the $M$
 posteriors we see that the Keck HC and VLT unexplained errors are
 somewhat closer to Normal than those for the Keck LC error group; the
 preference for low $M$ values here suggests that a fixed offset or
 bias would almost be sufficient to explain the residuals in this group.

In
 Figure \ref{fig3} we illustrate the corresponding posterior
 distributions of the pooled $\{\mu_\mathrm{sys}\}$ for each error
 group, which follow naturally our expectations given the posteriors of
 $\sigma_\mathrm{sys}$ and $M$.  The similarity in the distribution of
 $\{\mu_\mathrm{sys}\}$ between hypotheses (within each error group) is
 quite remarkable, however; the only exception being in the Keck LC
 error group where the dipole hypothesis contributes an effect
 indistinguishable from that of a
 bias term.  Finally, if we examine the posteriors of the hypothesis parameters for
the dipole shown in Figure \ref{fig4} we can confirm a fair agreement with
those of our biased, parametric models shown in Paper I.

\begin{figure*}
\vspace{0pc}
\begin{center}
\includegraphics[width=0.825 \textwidth]{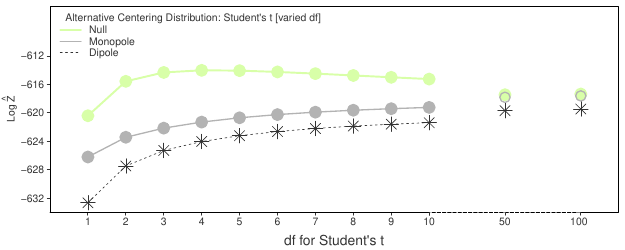} \end{center}
\vspace{-0.275cm}\caption{Investigation of prior-sensitivity in the
  marginal likelihoods of our three hypotheses for $\alpha$ variation
  in the quasar dataset.  Here we use importance sample reweighting of
our tempered likelihood posterior draws to explore the effect of
replacing the Normal centering distribution in our MDP model for the
unexplained error term with a Student's $t$ of varying degrees-of-freedom (df).
Lowering the df from 100 (near Normal) to 1 (Cauchy) increases the
Bayes factor in favour of the null.}
\label{fig5}
\end{figure*}

\subsection{Marginal Likelihoods}\label{mls}
The (log) marginal likelihoods for each hypothesis stated in Section \ref{hyp} under our semi-parametric error
model (after recursive estimation and
importance sample reweighting to convert from our conjugate prior proxies to
our true priors; cf.\ Section \ref{hdpm}) are as follows---null: $\log
\hat{Z} = -617.2$; monopole: $\log
\hat{Z} = -617.4$; and dipole: $\log
\hat{Z} = -619.2$ (with uncertainties $\Delta \log \hat{Z} \lesssim 0.1$ estimated via the [sample-based]
recursive asymptotic covariance matrix of \citealt{gil88}; and verified via repeat
simulation).  That is, we recover a posterior Bayes factor of
$\sim$$7.4$  in favour of the null over the dipole, which constitutes
weak support for the former under a uniform prior weighting for each
hypothesis; though, as we argue in Paper I, most cosmologists would in fact
presumably assign much greater prior weight to the null, concluding
that the existence of such a large-scale $\alpha$ dipole remains
exceedingly unlikely.  This Bayes factor 
ranking of hypotheses under our semi-parametric error model 
is reassuringly consistent with that reported for the biased
parametric error models in Paper I; the novelty being that our
use of a semi-parametric error model here has somewhat reduced the
potential for accusations of a subjective bias arising from the assumption
of a fixed functional form for the unexplained error term, which is
here allowed an ``infinitely flexible'',
non-parametric distribution via the MDP model.  The qualifier
``somewhat" in the above is important since, as we demonstrate in
Section \ref{psense} below, the MDP prior is not without its own
degree of prior-sensitivity.

Finally, it is
worth noting that the marginal likelihoods recovered (for all
hypotheses) under our Normal and skew Normal parametric error models in
Paper I were all greater than those recovered here.  Hence, although our
semi-parametric model has indeed offered an important robustness check
it has not performed better (in a BMS sense, under our stated priors) than our simplest
parametric models \textit{given the present dataset}.  With further
data though one might anticipate an increasing preference for the
(potentially ``more realistic'')
non-parametric family over these strict (``idealised'') parametric templates.

\subsection{Prior-Sensitivity Analysis}\label{psense}
In Figure \ref{fig5} we investigate the prior-sensitivity of the
marginal likelihoods for each of these three hypotheses under our
semi-parametric error model. In particular, we make use of the
Radon-Nikodym derivative for the DP \citep{dos09} and our importance
sample reweighting scheme (Section \ref{isr}) to explore the effect of
replacing the Normal centering distribution in our MDP model for the
unexplained error term with a Student's $t$ of varying degrees-of-freedom (df).
Lowering the df from 100 (near Normal) to 1 (Cauchy; ``fat-tailed'') increases the recovered
Bayes factor in favour of the null, and raises its absolute marginal
likelihood to a peak at a df of $4$, preserving the original BMS
rank-ordering of hypotheses under our nominal priors.  The computational time required
for the recomputation of these marginal likelihoods under the
alternative priors specified here was (as expected) indeed far less than that
required for the original parallel tempering scheme; confirming the
efficiency of the importance sample reweighting procedure for such
prior-sensitivity analyses. 

\section{Conclusions}\label{conclusions}
In this extension to our earlier work on the
Bayesian reanalysis of evidence for cosmic variation
in the fine structure constant we have developed a sophisticated
procedure for Bayesian model selection in the   
semi-parametric regime, allowing for efficient marginal likelihood estimation with prior-sensitivity
analysis.  Importantly, although the parameter space
described by the MDP form of our hierarchical prior features an
infinite-dimensional characterization we find that relatively trivial
modifications to the recursive algorithms presented in Paper I can allow
for a mode of computation directly analogous to the familiar case of strictly
parametric model selection.  Application of this methodology to the
problem at hand yields confirmation of our earlier finding in favour
of a null hypothesis for $\alpha$ variation coupled to a biased error
model for the instrumental subgroups of the Webb et al.\ team's quasar dataset; though
as before we do not anticipate the final resolution of this debate without
further data (such as that anticipated from the ongoing VLT Large
Program; 185.A-0745).

Finally, in a dedicated Section of this work we
have elucidated a number of 
 similarities between unexplained error problems in astronomy and
 clinical meta-analysis, which we hope may stimulate some  
 cross-disciplinary learning in this regard.

\section*{Acknowledgments}
\noindent\texttt{[1]} E.C.\ and A.N.P.\ are grateful for financial support from the Australian Research Council.

\label{lastpage}
\end{document}